# DYNAMICAL CONSTRAINTS ON THE FORMATION OF ELLIPTICAL GALAXIES

P.T. DE ZEEUW AND C.M. CAROLLO
*Sterrewacht Leiden*
*Postbus 9513, 2300 RA Leiden, The Netherlands*

**Abstract.** Recent work on the construction of spherical, axisymmetric and triaxial dynamical models for elliptical galaxies is reviewed briefly, including their role in providing evidence for dark halos and central black holes. The different orbital structures and shapes of low-mass and giant elliptical galaxies provide essential constraints on scenarios of galaxy formation.

## 1. Introduction

The aim of stellar dynamics is to reconcile the observed morphology and kinematics of galaxies, in order to determine their internal orbital structure. This provides information on the presence and importance of dark halos and massive central black holes, and on the connection between kinematics and stellar populations, i.e., between the motions and the physical properties of the stars.

Elliptical galaxies, as a class, are triaxial stellar systems with stationary or slowly tumbling figures (Binney 1976; Franx, Illingworth & de Zeeuw 1991). Their orbital structure depends on: (i) the rate of tumbling of their figure, (ii) their degree of triaxiality, and (iii) the presence and strength of a central concentration of mass (such as a stellar cusp or a black hole). Observations with the Hubble Space Telescope (HST) show that their luminosity distributions approach a power-law form $\rho(r) \propto r^{-\gamma}$ at small radii $r$, with $\gamma$ ranging between 0 and 2.5 (Crane et al. 1993; Ferrarese et al. 1994; Forbes, Franx & Illingworth 1995; Lauer et al. 1995; Carollo et al. 1995b). Thus, some galaxies have nearly-flat cores ($\gamma \sim 0$), while others have steep cusps ($\gamma > 1$; Figure 1).

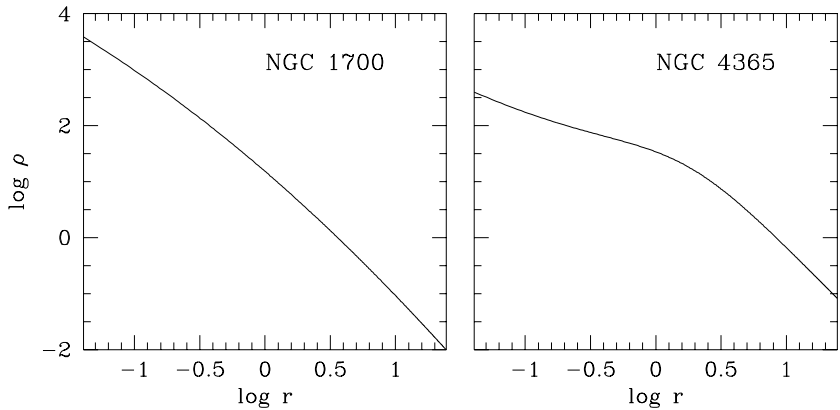

*Figure 1.* Two types of deprojected luminosity-density profiles (in $L_\odot/\mathrm{pc}^3$) for elliptical galaxies. The left panel shows the profile of the intermediate luminosity elliptical NGC 1700, which has a logarithmic slope $\gamma \approx 2$ down to the innermost measurable radii (given in arcsec). This galaxy has a steep cusp. The right panel shows the profile of the giant elliptical NGC 4365. It has a clear bend at $r_b \approx 2$ arcsec, inside which the profile is very shallow. Based on HST data from Carollo et al. (1995b). See also Forbes et al. (1995).

Theoretical dynamical modeling must answer the following key questions: (i) what is the full range of shapes, density profiles, and tumbling rates covered by equilibrium models, and (ii) what kind of internal velocity distributions do these have? The comparison of dynamical models with observations must subsequently establish which subset of all equilibrium models is occupied by elliptical galaxies. It is very likely that physical processes acting at formation, and/or subsequent dynamical evolution, have banished galaxies from certain parts of parameter space otherwise allowed by the simple requirement of stability. Delineating the region of permitted equilibria will provide essential information about scenarios of galaxy formation and mechanisms of galaxy evolution.

## 2. Spheres

An equilibrium model of a galaxy is fully specified by its phase-space distribution function (DF) $f(\vec{x}, \vec{v}) \geq 0$, which gives the density of stars at each position $\vec{x}$ and velocity $\vec{v}$. Jeans' theorem states that $f$ depends on $\vec{x}$ and $\vec{v}$ through the isolating integrals of motion admitted by the gravitational potential of the system. These integrals label the individual stellar orbits. Jeans' theorem therefore states that a dynamical model is defined once the number of stars that populate each orbit has been specified.

Jeans' theorem is most effective for spherical systems, in which the energy $E$ and the angular momentum $\vec{L}$ are conserved. Most models that have been constructed have DFs of the form $f(E, L^2)$, so that the velocity distribution is anisotropic, but does not have a preferred tangential direction. These models are particularly useful in studies of E0 galaxies. Important applications include the searches for dynamical evidence for massive dark halos and central black holes, i.e., for variations of the mass-to-light ratio $M/L$ with radius, in such systems. These studies require measurements of the entire line-of-sight velocity distribution — the velocity profile VP — as a function of radius, in order to distinguish variations in velocity anisotropy from variations in $M/L$ (e.g., Gerhard 1993; van der Marel 1994).

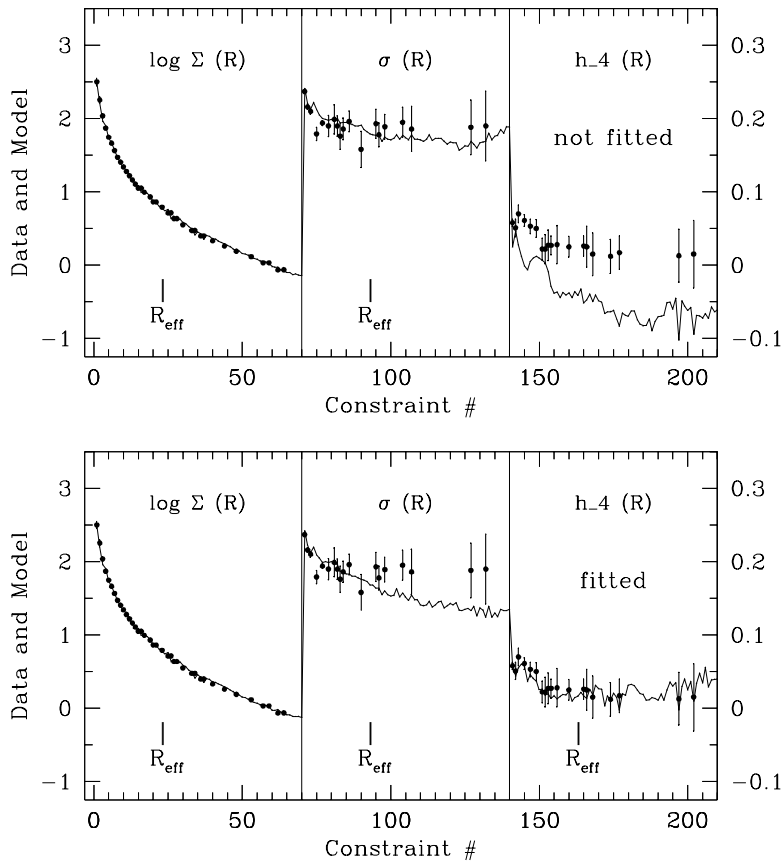

*Figure 2.* Models without dark halo for the E0 galaxy NGC 2434, compared with available photometry and kinematics. The upper panel shows the best fit to the radial profiles of the surface brightness $\Sigma$ and the velocity dispersion $\sigma$. The predicted values for the profile of the VP shape-parameter $h_4$ are shown as well, and are clearly inconsistent with the observed profile. The lower panel shows the best simultaneous fit to $\Sigma$, $\sigma$ and $h_4$. This again is unable to match the data, and in particular cannot reproduce the constant value of $\sigma$ out to large radii. NGC 2434 must therefore have a dark halo (Rix, de Zeeuw & Carollo 1995).

Carollo et al. (1995a) measured VP shapes out to about two half-light radii $R_e$ in the normal elliptical galaxy NGC 2434. The radial profile of the line-of-sight velocity dispersion $\sigma$ is flat to the outermost measured point. The VP shape parameter $h_4$, which indicates the lowest-order symmetric deviation from a Gaussian-shaped velocity profile, rules out strong tangential anisotropy. Based on a comparison with simple dynamical models, the authors argued that this E0 galaxy must have a dark halo. A similar result was obtained for another E0 galaxy, NGC 6703, by Jeske et al. (1995).

In order to demonstrate that a constant $M/L$ model can indeed be ruled out in NGC 2434, Rix, de Zeeuw & Carollo (1995) used a modified version of Schwarzschild's (1979) method for the construction of dynamical models. The surface brightness profile and the VP shapes are reproduced by populating individual orbits calculated in a chosen potential. Application to the NGC 2434 data showed that although the stellar orbits in a constant $M/L$ spherical model can reproduce the observed surface brightness and velocity dispersion profile, no such model can also fit the measured $h_4$-values (Figure 2). By contrast, inclusion of a plausible dark halo potential allows a simultaneous fit to all measurements. This is decisive stellar dynamical evidence for a dark halo surrounding a normal elliptical galaxy.

## 3. Axisymmetric Models

Despite our ignorance on the third integral $I_3$ in general axisymmetric potentials, knowledge of the two classical ones $E$ and $L_z$ has allowed investigation of axisymmetric dynamical models with DF $f(E, L_z)$. The actual calculation of such DFs has long been impeded by certain technical difficulties. Hunter & Qian (1993, hereafter HQ) developed a contour integral method that circumvents these difficulties, and allows calculation of $f(E, L_z)$ for a wide variety of finite and infinite mass models. When applied to the density $\rho(R, z)$ in a potential $\Psi(R, z)$, the method gives the unique $f_e(E, L_z)$ that is *even* in $L_z$ and that generates $\rho$. When applied to $R\rho\langle v_\phi\rangle(R, z)$, it gives the unique $f_o(E, L_z)$ that is *odd* in $L_z$ and generates the mean azimuthal streaming motions $\langle v_\phi\rangle$. It is not easy to obtain good observational data on the full two-dimensional mean line-of-sight velocity $\langle v_{\rm los}\rangle$ on the plane of the sky, from which the intrinsic azimuthal mean streaming field $\langle v_\phi\rangle(R, z)$ must be found. Therefore, a popular approach is to take the odd part $f_o$ as a product of the (positive) even part $f_e$ and a prescribed function such that $f = f_e + f_o$ is physical (non-negative).

Qian et al. (1995) applied the HQ method to high–quality kinematic data for M32 (van der Marel et al. 1994). No self-consistent $f(E, L_z)$ model can fit either the observed central peak in the line-of-sight velocity disper-sion $\sigma_{\rm los}$, or the steep central $\langle v_{\rm los}\rangle$ gradient. However, an $f(E, L_z)$ model with a central dark mass of $1.8 \times 10^6 M_\odot$ provides an astoundingly good fit to all available photometric and kinematic measurements (see Figure 11 in Qian et al. 1995; and Dehnen 1995). This does not prove that M32 has a large central black hole, since it remains to be demonstrated that a three-integral model without a central dark mass can be ruled out. However, the measured VP shapes show that in order to avoid a central dark mass, any model must combine a tangentially anisotropic velocity distribution outside 3 pc with a strongly radially anisotropic velocity distribution inside 3 pc. Even if a model with these properties and a DF $f \geq 0$ exists, it is not clear whether it would be stable, or, in fact, plausible. Cycle 5 HST/FOS observations have been collected to investigate this issue further. If, for the time being, we assume that the good fit of the simplest axisymmetric model with a central point mass is not a mere coincidence, then it is legitimate to ask: is the DF of M32 the end-product of stellar dynamical processes which have operated after its formation, possibly caused by the presence of a central dark mass? A positive answer would imply that such processes must have been effective in removing any dependence of the DF on a third integral of motion in less than the Hubble time.

Other recent applications of the HQ method include the construction of models with dark halos used to fit the extended kinematic absorption line measurements of elliptical galaxies (Carollo et al. 1995a), an analysis of the observable properties of small stellar disks embedded in the nuclei of ellipticals (van den Bosch & de Zeeuw 1995), and construction of models for the Galactic Bulge (Kuijken 1995; Hoogerwerf & Arnold 1995). In addition, following the HQ paper, a variety of other methods for generating $f(E, L_z)$-models have been developed by Dehnen & Gerhard (1994), Dehnen (1995), Kuijken (1995) and Magorrian (1995). It is safe to conclude that $f(E, L_z)$ axisymmetric models have now largely replaced spherical models as the standard theoretical template for a zeroth-order comparison with kinematic observations of flattened galaxies.

## 4. Triaxial Models

Triaxial potentials generally admit only one exact isolating integral, the orbital energy $E$. Although there are three planes of reflection symmetry, there are no symmetry axes, and no component of the angular momentum vector is conserved. Thus, two of the three integrals of motion are unknown. As a result, realistic models must be constructed by numerical means, such as Schwarzschild's method.

Since the orbital structure in a triaxial system depends on its central mass concentration, the internal dynamical structure of galaxies with flat cores differs from that of galaxies with steep cusps. Triaxial stellar systems with cores have a rich internal dynamical structure (Schwarzschild 1979; de Zeeuw 1985). There are four major orbit families. Two families circulate around the longest axis, and a third circulates around the shortest axis — these are called *tube* orbits; they carry all the angular momentum of the model. The fourth family is formed by the *box* orbits which have no net circulation and penetrate to the center of the model. Orbits of all four families must be occupied in any dynamical model, but many different orbit combinations, i.e., many different DFs, can reproduce the same triaxial density. Since there can be net streaming around both the long and the short axis, the total angular momentum vector (the 'rotation axis') can lie anywhere in the plane containing these two axes. The resulting projected kinematics can display a complex structure (Statler 1991, 1994; Arnold, de Zeeuw & Hunter 1994). Such complexity is indeed observed in the kinematics of giant elliptical galaxies (e.g., de Zeeuw & Franx 1991).

The three families of tube orbits are also present in models with cusps. Their mean streaming fields are therefore very similar to those of the models with cores. However, the few numerical constructions carried out to date indicate that whereas equilibrium models with cusp slopes $\gamma \lesssim 0.5$ probably can be built for all triaxial shapes, those with steeper inner profiles may exist only for near-axisymmetric shapes (Kuijken 1993; Schwarzschild 1993; Merritt & Fridman 1995). The reason is that in models with steep cusps or massive central black holes, the box orbits are replaced by minor orbit families and irregular orbits (Gerhard & Binney 1985; Merritt & Fridman 1995). These are associated with stable and unstable higher-order resonances between the oscillation frequencies along and perpendicular to the principal axes, and have been christened boxlets (Miralda–Escudé & Schwarzschild 1989). Boxlets display a large variety of shapes, but in substantially flattened triaxial models they cannot reproduce the characteristics of very elongated box orbits that remain close to the long axis. The latter are needed in any equilibrium model, owing to the fact that all tube orbits are elongated opposite to the figure of the model. These results are based on a small number of experiments, and need to be extended.

The fact that elliptical galaxies might have slowly tumbling figures has significant consequences for their orbital structure, since the Coriolis force distinguishes between direct and retrograde motion. Very little work has been done on the construction of tumbling triaxial galaxy models. The presence of a central density cusp or black hole is expected to influence the orbital structure less strongly than in a stationary triaxial system, because the Coriolis force causes box orbits to acquire net angular momentum and to become centrophobic (Vietri & Schwarzschild 1983). It is unknown whether a tumbling triaxial system can support a strong cusp.

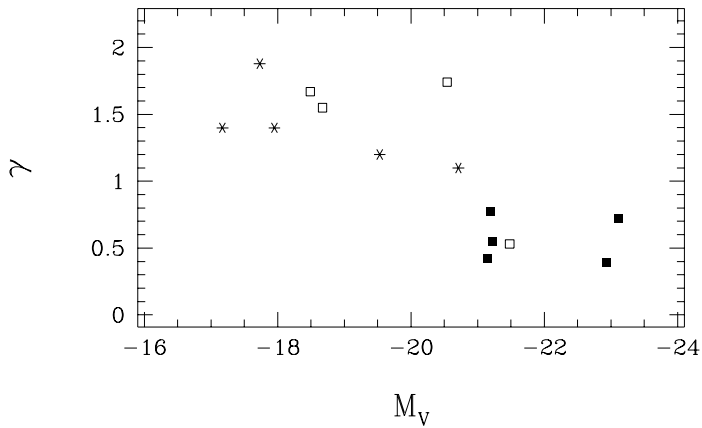

*Figure 3.* The inner logarithmic cusp slope of the deprojected density as a function of absolute $V$ magnitude, for a sample of elliptical galaxies. Different symbols indicate the quality of the deprojection (solid square = highest; open square = intermediate; stars = lowest; see Carollo et al. 1995b for details). Based on HST data from Lauer et al. (1995).

## 5. Implications and Conclusions

It is by now well established that the global properties of elliptical galaxies correlate with their mass. Large elliptical galaxies are red, have a high metal content, and have boxy isophotes. They rotate slowly, and are supported by anisotropic stellar velocity distributions. By contrast, low-luminosity ellipticals are bluer, are possibly less metal-rich, have disky isophotes, and rotate relatively fast (e.g., de Zeeuw & Franx 1991). Recent work with the HST has shown that also the nuclear properties, in particular the steepness of the central light profile, correlate with total luminosity (Figure 3).

When combined with the results on dynamical modeling of triaxial systems with cusps discussed in §4, these correlations suggest that (i) the giant ellipticals are indeed triaxial; (ii) the low-luminosity ellipticals are likely to have near-axisymmetric shapes. This difference in shape needs to be confirmed, and its origin needs to be investigated. It might have been imposed by the galaxy formation process, or might be the result of subsequent dynamical evolution, or both.

It is a pleasure to thank Hans-Walter Rix, Garth Illingworth, Marijn Franx, and Duncan Forbes, for permission to quote results prior to publication, and to thank Roeland van der Marel and Martin Schwarzschild for a careful reading of the manuscript. Part of this paper was written at the Institute for Advanced Study, with support from NSF Grant PHY 92–45317. CMC was supported by HCM grant ERBCHBICT940967 of the European Community.

**Discussion**

KOO: Given that luminous ellipticals seem to show the best evidence for massive black holes that can destroy the box orbits associated with triaxial shapes, how do you reconcile this with your conclusion that massive ellipticals are triaxial?

DE ZEEUW: I assume that the evidence you allude to is the fact that powerful radio sources are associated with giant ellipticals. Yet, as far as I am aware, we simply don't know whether all ellipticals have (quiescent) black holes, and in particular, whether the low-luminosity ones have them. The case of M32 is very interesting in this respect. As a result, it is hard to estimate the dynamical importance of these central point masses. Their effects may well be swamped by those induced by the central cusp.